\documentclass[manuscript,sigplan,authorversion,nonacm]{acmart}

\usepackage{xspace}
\usepackage{syntax}
\usepackage[section]{placeins}
\usepackage{subcaption}
\usepackage{graphicx}
\usepackage{caption}

\AtBeginDocument{%
  }

\setcopyright{acmcopyright}
\copyrightyear{2018}
\acmYear{2018}
\acmDOI{XXXXXXX.XXXXXXX}

\acmConference[Conference acronym 'XX]{Make sure to enter the correct
  conference title from your rights confirmation emai}{June 03--05,
  2018}{Woodstock, NY}
\acmPrice{15.00}
\acmISBN{978-1-4503-XXXX-X/18/06}

\newcommand{\eg}{\emph{e.g.,}\xspace}

\newcommand{\seclabel}[1]{\label{sec:#1}}
\newcommand{\figlabel}[1]{\label{fig:#1}}

\newcommand{\figref}[1]{Figure~\ref{fig:#1}}

\newcommand{\secref}[1]{Section~\ref{sec:#1}}

\usepackage{xcolor}
\usepackage{textcomp}
\usepackage{listings}
\usepackage{csquotes}
\usepackage{epigraph}
\usepackage{svg}

\definecolor{source}{gray}{0.9}
\lstdefinelanguage{scheme}{
	basicstyle=\ttfamily\small,
	comment=[l]{;},
	morecomment=[s]{\#|}{|\#},
	commentstyle=\color{purple}\ttfamily
}

\begin{document}

\title{$\mu\lambda\epsilon\delta$-Calculus:\\A Self Optimizing Language that Seems to Exhibit Paradoxical Transfinite Cognitive Capabilities}

\author{Ronie Salgado}
\email{ronie@ronie.cl}
\orcid{1234-5678-9012}
\affiliation{%
  \institution{Desromech EIRL - DCC, University of Chile}
  \city{Santiago}
  \country{Chile}
}

\begin{abstract}
Formal mathematics and computer science proofs are formalized using
Hilbert-Russell-style logical systems which are designed to not admit
paradoxes and self-refencing reasoning. These logical systems are natural way
to describe and reason syntactic about tree-like data structures. We found that
Wittgenstein-style logic is an alternate system whose propositional elements
are directed graphs (points and arrows) capable of performing paraconsistent
self-referencing reasoning without exploding. Imperative programming language are typically compiled
and optimized with SSA-based graphs whose most general representation is
the Sea of Node. By restricting the Sea of Nodes to only the data dependencies
nodes, we attempted to stablish syntactic-semantic correspondences with the
Lambda-calculus optimization. Surprisingly, when we tested our optimizer
of the lambda calculus we performed a natural extension onto the
$\mu\lambda$ which is always terminating. This always terminating algorithm is
an actual paradox whose resulting graphs are geometrical fractals, which
seem to be isomorphic to original source program. These fractal structures looks
like a perfect compressor of a program, which seem to resemble an actual
physical black-hole with a naked singularity. In addition to these surprising
results, we propose two additional extensions to the calculus to model the
cognitive process of self-aware beings: 1) $\epsilon$-expressions to model syntactic to semantic
expansion as a general model of macros; 2) $\delta$-functional expressions as
a minimal model of input and output. We provide detailed step-by-step
construction of our language interpreter, compiler and optimizer.
\end{abstract}

\ccsdesc[500]{Software and its engineering~Compilers}
\ccsdesc[500]{Software and its engineering~Interpreters}
\ccsdesc[500]{Software and its engineering~Translator writing systems and compiler generators}
\ccsdesc[500]{Software and its engineering~Dynamic compilers}
\ccsdesc[500]{Software and its engineering~Semantics}
\ccsdesc[500]{Software and its engineering~Syntax}
\ccsdesc[500]{Software and its engineering~Extensible languages}

\keywords{meta programming, lambda-calculus extension, paraconsistent logic, theory of computation}

\maketitle

\section{Introduction}
\epigraph{This book will perhaps only be understood by those who have already thought
the thoughs which are expressed in it or similar thoughts. It is therefore not
a text-book. Its object would be attained if it afforded to one who
read it with understanding. ... What can be said at all can be said clearly;
and whereof one cannot speak thereof one must be silent.}{Ludwig Wittgenstein}

In this article we describe the construction of an extension to the Lambda
calculus in terms of a contracting ordered directed multi-graph. The well
ordering of these graph data structures facilitates the usage of
transfinite induction since those graphs are isomorphic to well-ordered sets.
Formal semantics are normally constructed using a Russell\cite{whitehead1997principia} style theory of type,
which comes from the Hilbert formalization program. The problem of this logical
system is that they are designed so that paradoxes cannot exist. Instead we
propose using a Wittgenstein style logical system \cite{wittgenstein2023tractatus}
that allows paradoxical reasoning. In fact, we argue that four most
important impossibility in computer science are in-fact a limitation of the
underlying Russell style logical system. These four impossibility proofs are the
halting problem proof by Turing \cite{turing1936computable}, the Gödel
incompleteness theorem \cite{godel1992formally}, the Gentzen arithmetic
consistency proof, and the Tarski undefinability
proof \cite{tarski1933pojkecie}\cite{cieslinski2015tarski}.

These four impossibility proof are very similar between them, and the technique used is
always the Cantor diagonalization argument. A Wittgenstein style logical
system can represent diagonalization, and propositions which are constructed
with an infinite number of terms. The encoding of self-referencial paradoxes
in directed is possible by keeping the identity of the objects, and adding+
additional arrows. In fact, there is a very special object which can only be
encoded using this kind of systems: a graph with single vertex that points to
itself. This is the minimal meta-circular object definition. For reasons of
convenience, instead working directly with mutable directed graphs, we are using
only directed acyclic graphs. We encode this minimal circular object as $\mu x. x$.

This choice of encoding allows us to define our optimization algorithm as
directed graph rewriting system. This rewriting system uses memoization to
handle shared vertex, but it is also used for detecting and breaking 
self-referential cycles. The usage of directed graphs to facilitate sharing
has already been done in the context of functional programming
\cite{garner2012abstract} \cite{oliveira2012functional} \cite{shivers2005bottom}.
The usage of the greek letter $\mu$ to denote circular definitions, and graph
sharing has already been done before \cite{oliveira2012functional}. The lambda-mu
calculus it has already been formalized \cite{parigot1992lambdamu} \cite{laurent2004denotational}
and its strong normalization has already been proved \cite{david2009short}.
The novelty is on showing the connection with a restricted versions of the Sea
of Nodes for the lambda calculus, and the simplicity of a top-down memoized
implementation, which can be constructed by modifying a top-down interpreter.
Since we are using our optimizer for generating final values of programs, we do
not need to implement Global Code Motion \cite{click1995global}

In this article we are restricting ourself only to the purely functional lambda calculus plus some extensions for
modeling macros as syntactic expansion, and the sampling of external inputs to
model I/O feedback loops with an external universe. We are structuring this
step-by-step definition of our optimizing compiler using a similar mechanism
to the meta-circular evaluator of Reynolds \cite{reynolds1972definitional} In \secref{standard-lambda-calculus}
we provide a step-by-step construction on a lambda calculus interpreter in racket.
The reason for writing an interpreter is that we will use it as a base for
constructing our optimizing compiler in \seclabel{lambda-calculus-optimizer}.
Our optimizing compiler uses as Sea of Nodes IR \cite{click1995simple} which
is restricted to only data dependencies. We test our optimizing with several
instances that are known to explode \secref{optimizer-limit-testing}. Unlike
traditional programming language implementation, our optimization algorithm
is also our execution algorithm. In this article we are only listing the
essential elements of our proof of concept implementation. We provide
our complete source code under the MIT License in a GitHub
repository \cite{salgado-faila-mu-lambda-epsilon-delta}

\section{Standard Lambda Calculus Interpreter}\seclabel{standard-lambda-calculus}

\begin{figure*}[phbt]
\lstset{language=scheme}
\begin{lstlisting}
;; expr ::= <integer>
;;        | <boolean>
;;        | <identifier>
;;        | ()
;;        | (if <expr> <expr> <expr>)
;;        | (lambda (<identifier>*) <expr>)
;;        | (let ( (<identifier> <expr>)+ ) <expr>)
;;
;;        | (pair <expr> <expr>)
;;        | (first <expr>)
;;        | (second <expr>)
;;
;;        | (inject-left <expr>)
;;        | (inject-right <expr>)
;;        | (case <expr>)
;;
;;        | (<expr> <expr>)
;; Notes:
;; - Lambdas with multiple arguments are currified.
;; - Let expression with multiple arguments are normalized in a similar way to currified lambdas.
\end{lstlisting}
\caption{Syntax Grammar}
\figlabel{syntax-grammar}
\end{figure*}

\begin{figure*}[phbt]
\lstset{language=scheme}
\begin{lstlisting}
;; AST Syntactic nodes.
(struct stx-integer (value) #:transparent)
(struct stx-true () #:transparent)
(struct stx-false () #:transparent)
(struct stx-unit () #:transparent)
(struct stx-application (functional argument) #:transparent)
(struct stx-lambda (argument body) #:transparent)
(struct stx-let (name value body) #:transparent)
(struct stx-if (condition true-expression false-expression) #:transparent)
(struct stx-identifier (symbol) #:transparent)
(struct stx-pair (first second) #:transparent)
(struct stx-first (pair) #:transparent)
(struct stx-second (second) #:transparent)
(struct stx-inject-left (expression) #:transparent)
(struct stx-inject-right (expression) #:transparent)
(struct stx-case (expression left right) #:transparent)
\end{lstlisting}
\caption{Syntax Data Structures}
\figlabel{syntax-data-structures}
\end{figure*}

\begin{figure*}[phbt]
\lstset{language=scheme}
\begin{lstlisting}
;; parse-sexpr :: SExpression -> Syntax
(define (parse-sexpr sexpr)
  (cond
    [(integer? sexpr) (stx-integer sexpr)]
    [(boolean? sexpr) (if sexpr (stx-true) (stx-false))]
    [(symbol? sexpr) (stx-identifier sexpr)]
    [(list? sexpr) [match sexpr
      ([list] (stx-unit))
      ([list 'if condition true-expression false-expression]
        (stx-if (parse-sexpr condition) (parse-sexpr true-expression) (parse-sexpr false-expression)))
      ([list 'lambda arguments body] (foldr stx-lambda (parse-sexpr body) arguments))
      ([list 'let nameValues body]
        (foldr (lambda (nameValuePair body)
          (stx-let (first nameValuePair) (parse-sexpr (second nameValuePair)) body)
        ) (parse-sexpr body) nameValues))

      ;; Products
      ([list 'pair first second] (stx-pair (parse-sexpr first) (parse-sexpr second)))
      ([list 'first pair] (stx-first (parse-sexpr pair)))
      ([list 'second pair] (stx-second (parse-sexpr pair)))

      ;; Sums
      ([list 'inject-left expr] (stx-inject-left (parse-sexpr expr)))
      ([list 'inject-right expr] (stx-inject-right (parse-sexpr expr)))
      ([list 'case expr left right] (stx-case (parse-sexpr expr) (parse-sexpr left) (parse-sexpr right)))

      ;; Remaining case, applications.
      ([list-rest functional arguments] (foldl (lambda (a f) (stx-application f a)) (parse-sexpr functional) (map parse-sexpr arguments)))
    ]]
    [else (error "Unexpected syntax" sexpr)]))
\end{lstlisting}
\caption{Syntax Parser}
\figlabel{syntax-parser}
\end{figure*}

To implement an interpreter of a language, we need to first define its abstract
syntax grammar (See \figref{syntax-grammar}) and data structures for
representing the different syntactic elements (See \figref{syntax-data-structures}). We
also need a parser for this grammar, and since we are using Racket, a scheme dialect,
we can perform parse S-Expressions which are constructing by using the quote operator (See \figref{syntax-parser}).

\begin{figure*}[phbt]
\lstset{language=scheme}
\begin{lstlisting}
;; Environment :: Empty | Environment (Symbol -> Value)
(struct environment-empty () #:transparent)
(struct environment-child (parent symbol value) #:transparent)

;; lookup-valid-symbol :: Environment, Symbol -> Value. Error when not found.
(define (lookup-valid-symbol environment symbol)
  (match environment
    [(environment-empty) (error "Unbound symbol during interpretation: " symbol)]
    [(environment-child parent env-symbol value)
      (if (symbol=? env-symbol symbol)
        value
        (lookup-valid-symbol parent symbol))]
  ))
\end{lstlisting}
\caption{Environment Definition}
\figlabel{environment-definition}
\end{figure*}

Once we have a parsed abstract syntax tree (AST), we can start defining the
elements which are needed to implement an interpreter. In \figref{environment-definition}
we define the \emph{environment} or context used for performing a symbol lookup,
along with the symbol lookup function.

\begin{figure*}[phbt]
\lstset{language=scheme}
\begin{lstlisting}
;; Interpreter Value :: VInteger | VTrue | VFalse
;;                     | VClosure(Environment, Symbol, Syntax) 
;;                     | VPrimitive(Value -> Value)
;;                     | VUnit | VPair(Value, Value)
;;                     | VInjectLeft(Value) | VInjectRight(Value)
(struct val-integer (value) #:transparent)
(struct val-true () #:transparent)
(struct val-false () #:transparent)
(struct val-closure (environment argument body) #:transparent)
(struct val-pair (first second) #:transparent)
(struct val-primitive (implementation) #:transparent)
(struct val-unit () #:transparent)
(struct val-inject-left (value) #:transparent)
(struct val-inject-right (value) #:transparent)
\end{lstlisting}
\caption{Interpreter Value Definitions}
\figlabel{interp-value-definition}
\end{figure*}

In \figref{interp-value-definition} we define the allowed values of our
interpreter. The distinction between expansible syntactic values, versus
contractible semantic value is important. This distinction is the reason for not
reusing the AST structure for defining the values and its optimization. On the
next section we will continue expanding this of allowed values to define a first
version of the compiler.

\begin{figure*}[phbt]
\lstset{language=scheme}
\begin{lstlisting}
;; interp :: Environment, Syntax -> Value
(define (interp environment syntax)
  (match syntax
    [(stx-integer value) (val-integer value)]
    [(stx-true) (val-true)]
    [(stx-false) (val-false)]
    [(stx-unit) (val-unit)]
    [(stx-identifier name) (lookup-valid-symbol environment name)]
    [(stx-if condition true-expression false-expression)
      (define condition-value (interp environment condition))
      (match condition-value
        [(val-true) (interp environment true-expression)]
        [(val-false) (interp environment false-expression)])]
    [(stx-application functional argument)
      (define functional-value (interp environment functional))
      (define argument-value (interp environment argument))
      (interp-apply-value-with-value environment functional-value argument-value)]
    [(stx-lambda argument body)
      (val-closure environment argument body)]
    [(stx-let name value body)
      (define child-env (environment-child environment name (interp environment value)))
      (interp child-env body)]

    ;; Products
    [(stx-pair first second)
      (define first-value (interp environment first))
      (define second-value (interp environment second))
      (val-pair first-value second-value)]
    [(stx-first pair)
      (match (interp environment pair)
        [(val-pair first second) first])]
    [(stx-second pair)
      (match (interp environment pair)
        [(val-pair first second) second])]

    ;; Sums
    [(stx-inject-left expression) (val-inject-left (interp environment expression))]
    [(stx-inject-right expression) (val-inject-right (interp environment expression))]
    [(stx-case expression left-case right-case)
      (match (interp environment expression)
        [(val-inject-left injected-value) (interp-apply-with-value environment left-case injected-value)]
        [(val-inject-right injected-value) (interp-apply-with-value environment right-case injected-value)])]
  ))
\end{lstlisting}
\caption{Interpreter Definition}
\figlabel{interp-definition}
\end{figure*}

The full definition code for our interpreter is given in \figref{interp-definition}.
This is a standard recursive implementation of the lambda-caculus defined by scheme.
On the next section we will start converting this implementation into compiler. 

\section{Lambda Calculus Optimizing Compiler}\seclabel{lambda-calculus-optimizer}

\begin{figure*}[phbt]
\lstset{language=scheme}
\begin{lstlisting}
;; Compiler Value ::
;;                 | VApply(Value, Value)
;;                 | VArgument
;;                 | VIf(Value, Value, Value) 
;;                 | VInteger | VTrue | VFalse
;;                 | VClosure(Environment, Symbol, Syntax) 
;;                 | VPrimitive(Value -> Value)
;;                 | VUnit
;;                 | VPair(Value, Value)
;;                 | VFirst(Pair)
;;                 | VSecond(Pair)
;;                 | VInjectLeft(Value)
;;                 | VInjectRight(Value)
;;                 | VCase(Value, Value, Value)
(struct val-apply (functional argument) #:transparent)
(struct val-argument () #:transparent)
(struct val-if (condition true-block false-block) #:transparent)
(struct val-integer (value) #:transparent)
(struct val-true () #:transparent)
(struct val-false () #:transparent)
(struct val-lambda (argument body) #:transparent)
(struct val-pair (first second) #:transparent)
(struct val-first (pair) #:transparent)
(struct val-second (pair) #:transparent)
(struct val-primitive (implementation) #:transparent)
(struct val-unit () #:transparent)
(struct val-inject-left (value) #:transparent)
(struct val-inject-right (value) #:transparent)
(struct val-case (value left-case right-case) #:transparent)
\end{lstlisting}
\caption{Compiler Value Definitions}
\figlabel{compiler-value-definitions}
\end{figure*}

\begin{figure*}[phbt]
\lstset{language=scheme}
\begin{lstlisting}
;; comp :: Environment, Syntax -> Value
(define (comp environment syntax)
  (match syntax
    [(stx-integer value) (val-integer value)]
    [(stx-true) (val-true)]
    [(stx-false) (val-false)]
    [(stx-unit) (val-unit)]
    [(stx-identifier name) (lookup-valid-symbol environment name)]
    [(stx-if condition true-expression false-expression)
      (define condition-value (comp environment condition))
      (define true-value (comp environment true-expression))
      (define false-value (comp environment false-expression))
      (val-if condition-value true-value false-value)]
    [(stx-application functional argument)
      (define functional-value (comp environment functional))
      (define argument-value (comp environment argument))
      (val-apply functional-value argument-value)]
    [(stx-lambda argument body)
      (define argument-value (val-argument))
      (define closure-environment (environment-child environment argument argument-value))
      (define closure-body (comp closure-environment body))
      (val-lambda argument-value closure-body)]
    [(stx-let name value body)
      (define child-env (environment-child environment name (comp environment value)))
      (comp child-env body)]

    ;; Products
    [(stx-pair first second)
      (define first-value (comp environment first))
      (define second-value (comp environment second))
      (val-pair first-value second-value)]
    [(stx-first pair) (val-first (comp environment pair))]
    [(stx-second pair) (val-second (comp environment pair))]

    ;; Sums
    [(stx-inject-left expression) (val-inject-left (comp environment expression))]
    [(stx-inject-right expression) (val-inject-right (comp environment expression))]
    [(stx-case expression left-case right-case)
      (define expression-value (comp environment expression))
      (define left-case-value (comp environment left-case))
      (define right-case-value (comp environment right-case))
      (val-case expression-value left-case-value right-case-value)]
  ))
\end{lstlisting}
\caption{Compiler Definition}
\figlabel{compiler-definitions}
\end{figure*}

\begin{figure*}[phbt]
\lstset{language=scheme}
\begin{lstlisting}
;; val-recurse-children :: DagContext, Value -> Value
(define (val-recurse-children value rec)
  (match value
    [(val-apply functional argument) (val-apply (rec functional) (rec argument))]
    [(val-argument) value]
    [(val-if condition true-block false-block) (val-if (rec condition) (rec true-block) (rec false-block))]
    [(val-integer _) value]
    [(val-true) value]
    [(val-false) value]
    [(val-lambda argument body) (val-lambda (rec argument) (rec body))]
    [(val-pair first second) (val-pair (rec first) (rec second))]
    [(val-first pair) (val-first (rec pair))]
    [(val-second pair) (val-second (rec pair))]
    [(val-primitive _) value]
    [(val-inject-left value) (val-inject-left (rec value))]
    [(val-inject-right value) (val-inject-right (rec value))]
    [(val-case value left-case right-case) (val-case (rec value) (rec left-case) (rec right-case))]))
\end{lstlisting}
\caption{Value Recursor}
\figlabel{value-recursor}
\end{figure*}

To turn our interpreter into a compiler, we need to add additional values
for forms such as \emph{if}. For the full definition of these values
see \figref{compiler-value-definitions}. Each one of these values has a
correspondence with a data only node in a Sea of Nodes IR \cite{click1995simple}.
For the full definition of our compiler from syntax into IR see \figref{compiler-definitions}.

\begin{figure*}[phbt]
\lstset{language=scheme}
\begin{lstlisting}
;; dag-memoize :: DagContext, (Unit -> Any)
;; Mutable store used for graph transform algorithm. Keys are compared by identity.
(define (dag-memoize context function-name value transform)
  (define memoization-table (dag-get-memoization-table context function-name))
  (if (hash-has-key? memoization-table value)
    (match (hash-ref memoization-table value)
      [(dag-pending-token) (error "Cyclic expansion of " value)]
      [memoized-result memoized-result])
    (begin
      (hash-set! memoization-table value (dag-pending-token))
      (let [(transform-result (transform))]
        (hash-set! memoization-table value transform-result)
        transform-result))))
\end{lstlisting}
\caption{DAG Memoization}
\figlabel{dag-memoization}
\end{figure*}

Optimizations are done by applying a set of reduction rules. To facilitate
writing these reductions, we made a recursor (See \figref{value-recursor}) that
encapsulates the concept of perform a DAG top-down traversal and rewriting. To
facilitate memoization we made an utility where a memoization cache is used to
keep the identity of shared graph nodes (See \figref{dag-memoization}). The
internal implementation of the memoization requires actual hash tables for
O(1) amortized memoization lookup. In a purely functional programming language
this hash table can only be implemented as binary search tree whose acceses
are O(log n).

\begin{figure*}[phbt]
\lstset{language=scheme}
\begin{lstlisting}
;; reduction-rule :: DagContext, Value -> Value
(define (reduction-rule context value)
  (match value
    [(val-apply (val-primitive primitive) argument) #:when (is-constant-val context argument)
      (primitive argument)] ;; Evaluate primitive with constants.
    [(val-apply (val-lambda argument-definition body) argument-value)
      (substitute (dag-context) argument-definition argument-value body)]
    [(val-if (val-true) true-block _) true-block]
    [(val-if (val-false) _ false-block) false-block]
    [(val-first (val-pair first _)) first]
    [(val-second (val-pair _ second)) second]
    [(val-case (val-inject-left value) left-case _) (val-apply left-case value)]
    [(val-case (val-inject-right value) _ right-case) (val-apply right-case value)]
    [value value]))

;; reduce-once :: DagContext, Value -> Value
(define (reduce-once context value)
  (dag-memoize context 'reduce-once value (lambda ()
      (define with-reduced-child (val-recurse-children value (lambda (child) (reduce-once context child))))
      (reduction-rule context with-reduced-child))))

;; reduce :: DagContext, Value -> Value
;; Reduce until achieving a fixpoint.
(define (reduce context value)
  (dag-memoize context 'reduce value (lambda ()
      (define with-reduced-child (val-recurse-children value (lambda (child) (reduce context child))))
      (define reduced-once (reduction-rule context with-reduced-child))
      (if (eq? with-reduced-child reduced-once)
        reduced-once
        (reduce context reduced-once)))))
\end{lstlisting}
\caption{Reduction Rules and Methods}
\figlabel{reduction-rule-and-methods}
\end{figure*}

The reduction rules and methods are described in \figref{reduction-rule-and-methods}.
All of the reduction rules are described in the aptly named \emph{reduction-rule} function.
These reduction rules can be applied in two different ways: 1) as a single reduction
step (\eg reduce once); and 2) as an iterative reduction applications until a fixed point is reached.

\begin{figure}[phbt]
\centering

\begin{subfigure}[b]{0.16\textwidth}
\centering
\includesvg[width=6cm]{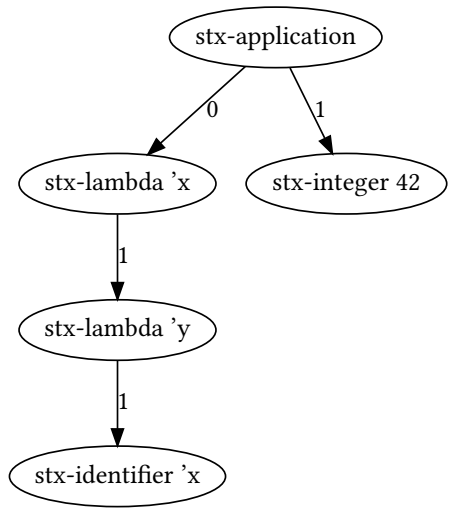}
\caption{Syntax Tree}
\end{subfigure}
\quad

\begin{subfigure}[b]{0.16\textwidth}
\centering
\includesvg[width=6cm]{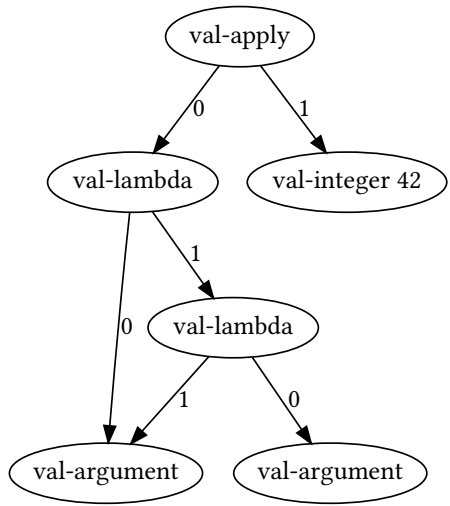}
\caption{Compilation DAG}
\end{subfigure}
\quad
    
\begin{subfigure}[b]{0.16\textwidth}
\centering
\includesvg[width=6cm]{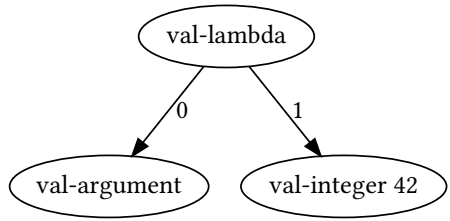}
\caption{Reduced Once DAG}
\end{subfigure}
\quad

\begin{subfigure}[b]{0.16\textwidth}
\centering
\includesvg[width=6cm]{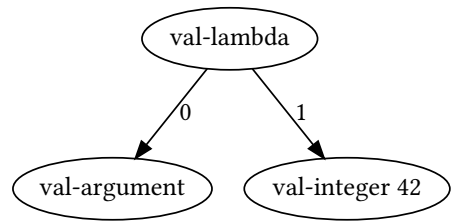}
\caption{Reduced DAG}
\end{subfigure}

\caption{Compilation process for '((lambda (x y) x) 42)}
\figlabel{lambda-xy-compilation}
\end{figure}

In \figref{lambda-xy-compilation} we have a simple example of compilation and
optimization of a simple expression ($(\lambda x y. x) 42$) that gets
reduced to a simpler along the different compilation and optimization phases. The
end result is a DAG for a curried version of the expression ($\lambda y. 42$).

\section{Optimizing Compiler Limit Testing}\seclabel{optimizer-limit-testing}

\begin{figure}[htb]
\centering

\begin{subfigure}[b]{0.16\textwidth}
\centering
\includesvg[width=10cm]{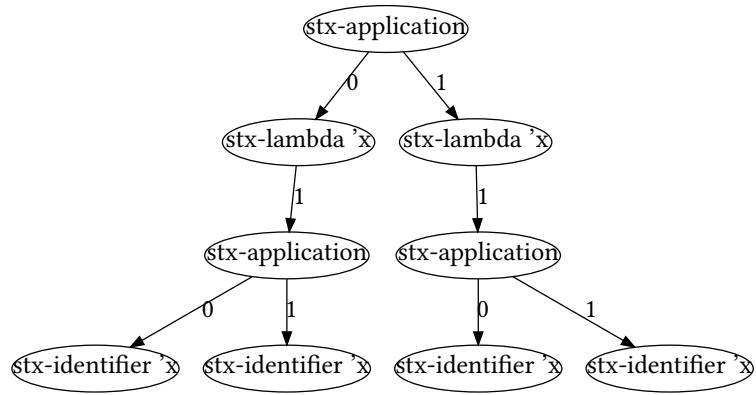}
\caption{Syntax Tree}
\end{subfigure}
\quad

\begin{subfigure}[b]{0.16\textwidth}
\centering
\includesvg[width=6cm]{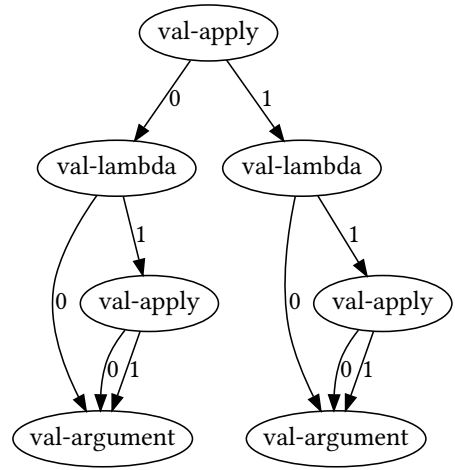}
\caption{Compilation DAG}
\end{subfigure}
\quad
    
\begin{subfigure}[b]{0.16\textwidth}
\centering
\includesvg[width=4cm]{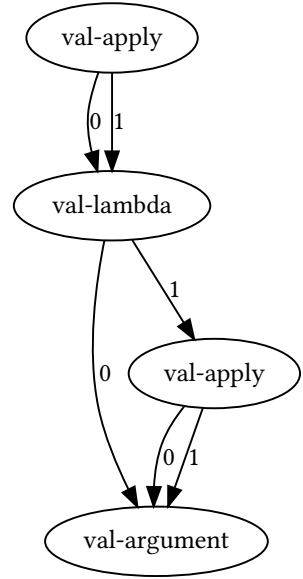}
\caption{Reduced Once DAG}
\end{subfigure}
\quad

\caption{Compilation process for Omega}
\figlabel{omega-compilation-ver1}
\end{figure}

A more interesting example is the compilation of the Omega combinator
($\lambda x. xx$)($\lambda x.xx$) which corresponds to an infinite loop. The
omega combinator is equivalent to while true loop in an imperative language
such as C. In \figref{omega-compilation-ver1} we see the compilation and
optimization process for the omega combinator. Here we can see that the usage of
a graph based representation allows to specify a function that is applied to
itself. Unfortunately, this version of the compiler cannot optimize completely
omega because is not detecting properly the fixed point element. To solve that
we had to introduce unification onto the optimizer. For unification we have a hash-table
from a semantic object into a representative already optimized version of that semantic
object. This unification process is equivalent to the well known union-find
data structure. The union-find data structure has a complexity of the order of
the reciprocal of the Ackermann Function which is a extremely slow growing
function. In this case, these extremely slow growing function seems to be the
actual complexity of our optimization algorithm, and since this is with arbitrary
turing complete programs, then we suspect that this actually the key for
achieving transfinite computation. However, an extra ingredient is needed for
performing transfinite induction, which is the limiting case. The unification
source code listing is in \figref{unification-code}.

The expression ($\lambda x. xxx$)($\lambda x.xxx$) is a well known expansive
non-terminating lambda calculus expression. When we tried the full reduction with
this expression after implementing unification, instead of getting a
non-terminating optimization, we obtained an error because we needed to
reference a previously seen reduction. This is the limiting case required for
the transfinite induction algorithm. We solved this case by marking in a table
a token for pending reduction. If this pending reduction is referenced, instead
of reducing it again, we return a $\mu$ argument expression, and we mark the
cyclic case. For the partial source code listing, see \figref{limiting-case-code}
for the scaffolding, see \figref{cyclic-reduction-rules} for the updated reduction function,
and for the reduction of this expansive expression see \figref{expansive-omega3-compilation}
and \figref{expansive-omega3-reduced} for the final self-referential reduced version.

\begin{figure*}[phbt]
\lstset{language=scheme}
\begin{lstlisting}
;; can-unify :: DagContext, Value -> Value
(define (can-unify value)
  (match value
    [(val-argument) #f]
    [(val-muargument) #f]
    [_ #t]))

;; unify :: DagContext, Value -> Value
(define (unify context value)
  (define (do-unify)
    (define unification-value (dag-unif-val value))
    (define unification-table (dag-get-memoization-equal-table context 'unification))
    (if (hash-has-key? unification-table unification-value)
      (hash-ref unification-table unification-value)
      (begin
        (hash-set! unification-table unification-value value)
        (let ([with-unified-children (val-recurse-children value (lambda (child) (unify context child)))])
          (hash-set! unification-table unification-value with-unified-children)
          value)
      )))

  (if (can-unify value)
    (do-unify)
    value))
\end{lstlisting}
\caption{Methods used for unification}
\figlabel{unification-code}
\end{figure*}

\begin{figure*}[phbt]
\lstset{language=scheme}
\begin{lstlisting}
;; DagContext :: Dict(Symbol, Dict(Any, Any))
;; Mutable store used for graph transform algorithm. Keys are compared by identity.
(define dag-context make-hash)
(struct dag-pending-token ())
(struct dag-pending-cylic-token (muarg is-cyclic))

;; dag-cyclic-memoize :: DagContext, (Unit -> Any)
;; Common scaffolding required for cyclic-reducing dag rewriting algorithms.
;; This seems to be related to transfinite induction/recursion, and self-referencing systems.
;; Needed for the reduction of ((lambda (x) (x x x)) (lambda (x) (x x x))).
(define (dag-cyclic-memoize context function-name value muarg-function mu-function transform)
  (define memoization-table (dag-get-memoization-table context function-name))
  (if (hash-has-key? memoization-table value)
    (match (hash-ref memoization-table value)
      [(dag-pending-cylic-token muarg is-cyclic-box) (begin
        (set-box! is-cyclic-box \#t)
        muarg)]
      [memoized-result memoized-result])
    (begin (let [(muarg (muarg-function)) (is-cyclic-box (box \#f))]
      (hash-set! memoization-table value (dag-pending-cylic-token muarg is-cyclic-box))
      (let [(transform-result (transform))]
        (let [(result (if (unbox is-cyclic-box)
            (mu-function muarg transform-result)
            transform-result
          ))]
          (begin
            (hash-set! memoization-table value result)
            result
          )))))))
\end{lstlisting}
\caption{Scaffolding for handling cyclic cases}
\figlabel{limiting-case-code}
\end{figure*}

\begin{figure*}[phbt]
\lstset{language=scheme}
\begin{lstlisting}
;; reduction-rule :: DagContext, Value -> Value
(define (reduction-rule context value)
  (match value
    [(val-apply (val-primitive primitive) argument) #:when (is-constant-val context argument)
      (primitive argument)] ;; Evaluate primitive with constants.
    [(val-apply (val-lambda argument-definition body) argument-value)
      (substitute context argument-definition argument-value body)]
    [(val-mu argument body)
      (if (uses-var? context body argument)
        value
        body)]
    [(val-if (val-true) true-block _) true-block]
    [(val-if (val-false) _ false-block) false-block]
    [(val-if expr result result) result]
    [(val-first (val-pair first _)) first]
    [(val-second (val-pair _ second)) second]
    [(val-case (val-inject-left value) left-case _) (val-apply left-case value)]
    [(val-case (val-inject-right value) _ right-case) (val-apply right-case value)]
    [value value]))

;; reduce-once :: DagContext, Value -> Value
(define (reduce-once context value)
  (dag-cyclic-memoize context 'reduce-once value val-muargument val-mu (lambda ()
      (define with-reduced-child (unify context (val-recurse-children value (lambda (child) (reduce-once context child)))))
      (unify context (reduction-rule context with-reduced-child)))))

;; reduce :: DagContext, Value -> Value
;; Reduce until achieving a fixpoint.
(define (reduce context value)
  (dag-cyclic-memoize context 'reduce value val-muargument val-mu (lambda ()
      (define with-reduced-child (unify context (val-recurse-children value (lambda (child) (reduce context child)))))
      (define reduced-once (unify context (reduction-rule context with-reduced-child)))
      (if (eq? with-reduced-child reduced-once)
        reduced-once
        (reduce context reduced-once)))))
\end{lstlisting}
\caption{Cyclic reduction rules}
\figlabel{cyclic-reduction-rules}
\end{figure*}

\begin{figure}[htb]
\centering

\begin{subfigure}[b]{0.16\textwidth}
\centering
\includesvg[width=4cm]{omega3-comp}
\caption{Compilation DAG}
\end{subfigure}
\quad
    
\begin{subfigure}[b]{0.16\textwidth}
\centering
\includesvg[width=4cm]{omega3-reduced-once}
\caption{Reduced Once DAG}
\end{subfigure}
\quad

\begin{subfigure}[b]{0.16\textwidth}
\centering
\includesvg[width=4cm]{omega3-reduced-once2}
\caption{Reduced Twice DAG}
\end{subfigure}
\quad

\caption{Compilation process for Expansive Omega}
\figlabel{expansive-omega3-compilation}
\end{figure}

\begin{figure}[htb]
\centering
\includesvg[width=4cm]{omega3-reduced}
\caption{Self referential reduction result}
\figlabel{expansive-omega3-reduced}
\end{figure}

\section{Epsilon Delta Extensions}\seclabel{future-work}

\paragraph{Macros via Epsilon Functional Expansion} As a first extension to our
system we are adding the notion that a syntactic element can be expanded into
a semantic on a given context. In other words, the whole compilation
from syntax values to semantic values can be seen as macro expansion. With this
scheme of language definition, we define a macro as function from contextual
syntax to contextless semantics. For introducing this macros, added an $\epsilon$
functional definition which is analog to the $\lambda$ functional. Another
requirement for supporting macros is the need for exposing and manipulating the
lookup environment as object and values in the target environment. In our current
proof of concept implementation we are facing some issues with the expansion
of environment lookups, and for this reason we are not providing a full listing
of this part of the calculus.

\paragraph{External World Communication via Delta functionals} The $\mu\lambda$-calculus
is complete and consistent theories whose program have a correspondence with an
ordinal number. By covering all of the ordinal numbers, we know that the
cardinality of the computable space is the same one as the natural numbers. We
know that they are programs an values which are actual real numbers coming from
the outside world. All of the communications between the inner and outer world
are made through the passage of contextual syntactic elements. If we want to
operate with real numbers, we have to gather them by sampling the outer world.
For this reason, we propose extending the calculus with a $\delta$ functional
whose job is to denote I/O ports with the exterior world. An inner delta
functional can be replaced directly by a $\lambda$ expression. These functional
objects are equivalent. 

\paragraph{Computational Limits} We choose to name these two extension
$\epsilon$ and $\delta$ to have an analogy with concept of limits. In fact, we
are providing these two elements with the explicit objective of escaping the
boundaries of closed Turing complete systems, into the boundary of open systems.

\section{Mathematical, Physical and Philosophical Interpretations}\seclabel{future-work}

\paragraph{Mathematical Interpretations}We implemented an optimizer using
directed acyclic graphs, which are isomorphic
to directed cyclic graphs via de $\mu$ expressions. This optimizer always
terminates, and it is capable of implementing and running the Lambda calculus
until a normal form is achieved. All of the reduction rules and graph elements
are completely ordered. This ordering means that this particular variant of the
lambda-mu calculus is a well-ordered set. Well-ordered sets have a
correspondence with transfinite ordinal numbers, and they support transfinite
induction and recursion. This is the first time that we see transfinite
computation inside existing Turing complete computation. The key for this is
noticing that programming language have ordered elements, and these elements
have an identity which typically corresponds to the pointer address of an object.

\paragraph{Physical Interpretations} Transfinite computation poses a serious
challenge to the causality principle. In fact, in our opinion this principle is
a complete illusion caused by a self-aware being. In fact, we propose that we
can describe some self-aware entity using something similar to this
expression: $\mu x.\delta y$. The top-down reduction process can be seen as
the entanglement of computation. An iteration of a feedback loop can be seen as
taking a picture and observing a particular state of affairs. In our opinion,
this algorithm is an actual logical paradox whose singularity is completely naked.
In other words, the $\mu$ functional is a constructor of singularities which can
be seen as black holes, white holes, and even as a cellular membrane. The
smallest and greatest black-hole is denoted by $mu x. x$, which corresponds
to a single self-referencing vertex in a directed graph. The liar paradox the
following direct representation: $mu x. ~x$. The halting problem can also be
studied by finding fixed points using these black-holes that represent
transfinite computation. These self-referential cyclic structures can be seen
as fractal geometry, and the resulting optimized version is the minimal fractal
program that is isomorphic to the original source program.

\paragraph{Related Physical Interpretations} Wolfram has proposed the usage of
hypergraph rewriting systems as a model for a theory of everything in \cite{wolfram2020class}.
Connections with general relativity \cite{gorard2020some} and quantum mechanics
\cite{gorard2020somequantum} have already been done with the Wolfram automata.
The mass-energy equivalence principle is well known from Einstein in Physics. In
recent years, a mass-energy-information equivalence principle has been
proposed \cite{vopson2019mass}\cite{vopson2022experimental}.

\paragraph{Philosophical Interpretations} For us the philosophical interpretation
is the most nuanced part of this algorithm. This algorithm can be seen as playing
a game, until getting bored of that game because you notice it is exactly the
same as a previous one. This algorithm is a true singularity, and it is also
exhibiting self-encrypting properties. This means that the content of this very
same article can be read by two kind of living being: 1) beings that are
constantly improving their self-awareness, that knows some bits from physics,
psychology and computer science; 2) the second kind of beings that can read and
understand completely this article are \emph{super-intelligent} beings.

\paragraph{Super Intelligence thought experiment} When we wrote the tentative
equation for a self-aware living being ($\mu x.\delta y$), we also did a thought
experiment. What can happen if this equation is programmed in a virtual reality
environment? Would it converge into mirror of myself? would transform myself
onto a super intelligent being? The biggest issue with this kind of thought
experiment is that since causality principle was already broken. In other words,
by just thinking on the experiment some results can be seen before even
starting to implement them. This is the paradoxical power of transfinite
computation. Transfinite computation has already been discussed in terms of
hypothetical relativistic machines \emph{welch2007turing}. Since the principle
of causality is broken at this level, then the super-intelligent human brain has
to become a new agent of causality by balancing the flows of information between
its inner psychological world, and its outer physical world.

\section{Future Work}\seclabel{future-work}
In the future we would like to extend this approach by using the control flow
nodes from the full Sea of Nodes IR. We are planning on using these tools for
implementing highly optimized object-functional programming languages. We are
thinking on experimenting with the construction of a Smalltalk style language
using the Sea of Nodes as IR. We are interested on testing the virtual reality
hypothesis. And we would like to test this system for constructing robots
without having to resort to large language models and clusters of GPU.

\section{Conclusions}\seclabel{conclusions}

\paragraph{Acknowledgment} We acknowledge receiving the idea of implementing a
restricted sea of nodes from professor Éric Tanter, and professor Matías Toro.
Without them, we might not even explored this section of the problem space.

\paragraph{A New Frontier} We wanted to show how a restricted version Sea of
Nodes has a direct correspondence to the lambda calculus. Our initial objective
was to have a basic tutorial style description on how to work with graph based
IR. We never expected to encounter a rich and vast algorithm that allows
work with some infinite problems. We will continue exploring these venue of
transfinite induction and recursion. We will focus on the geometrical properly
and in finding better ways to visualize and manipulate the fractals that give
rise to this computational model.

\bibliographystyle{ACM-Reference-Format}
\bibliography{references}

\end{document}